\documentclass{article}

\usepackage{amsmath}
\usepackage{caption}
\usepackage{arxiv}
\usepackage{graphicx}
\usepackage[utf8]{inputenc} 
\usepackage[T1]{fontenc}    
\usepackage{hyperref}       
\usepackage{url}            
\usepackage{booktabs}       
\usepackage{amsfonts}       
\usepackage{nicefrac}       
\usepackage{microtype}      
\usepackage{lipsum}
\usepackage{multicol}
\usepackage{multirow}
\usepackage{subcaption}
\usepackage{arxiv}
\usepackage[square, numbers, comma, sort&compress]{natbib}

\title{A low-cost real-time 3D imaging system for contactless asthma observation}

\author{Sheona M.M.D.P. Sequeira, Beril Sirmacek \\
  Robotics and Mechatronics\\
  University of Twente\\
  Enschede, The Netherlands \\
  \texttt{b.sirmacek@utwente.nl} \\
}

\begin{document}
\maketitle

\begin{abstract}
Asthma is becoming a very serious problem with every passing day, especially in children. However, it is very difficult to detect this disorder in them, since the breathing motion of children tends to change when they reach an age of 6. This, thus makes it very difficult to monitor their respiratory state easily. In this paper, we present a cheap non-contact alternative to the current methods that are available. This is using a stereo camera, that captures a video of the patient breathing at a frame rate of 30Hz. For further processing, the captured video has to be rectified and converted into a point cloud. The obtained point clouds need to be aligned in order to have the output with respect to a common plane. They are then converted into a surface mesh. The depth is further estimated by subtracting every point cloud from the reference point cloud (the first frame). The output data, however, when plotted with respect to real time produces a very noisy plot. This is filtered by determining the signal frequency by taking the Fast Fourier Transform of the breathing signal. The system was tested under 4 different breathing conditions: deep, shallow and normal breathing and while coughing. On its success, it was tested with mixed breathing (combination of normal and shallow breathing) and was lastly compared with the output of the expensive 3dMD system. The comparison showed that using the stereo camera, we can reach to similar sensitivity for respiratory motion observation. The experimental results show that, the proposed method provides a major step towards development of low-cost home-based observation systems for asthma patients and care-givers.
\end{abstract}


\section{Introduction and Background}
The need for monitoring respiratory motion is increasing rapidly. There are a lot of people that tend to suffer from breathing disorders, asthma being the prime one. Asthma is a very serious condition in which it becomes extremely difficult to breathe, due to the airway passage becoming very narrow and swollen. This, thus prevents the proper flow of air into the lungs. This disorder, in turn results in the patient getting extremely tired fast, making it difficult for them to perform daily activities that a normal individual would do. Asthma is sadly a condition which cannot be healed permanently, but the intensity of the asthmatic attacks can be reduced by taking proper treatment and care. However, this condition is extremely difficult to be properly diagnosed in children \cite{ref1}. This is due to the fact that children who show recurring symptoms of asthma at ages below 6 not necessarily will continue to suffer from asthma later on in their lives. Thus, this usually tends to go undiagnosed with the hope of it being a passing phase. While on one hand the children with mild asthma-like symptoms might no longer suffer from it after the age of 6 or more, children with more severe asthmatic conditions often tend to still suffer. For this very purpose, monitoring the respiratory motion especially in children is vital.

There are many devices that can be used for tracking respiratory motion \cite{ref2}. These devices might be further categorized based on the motion of the body, that is, they can be invariant to rigid body motion or not. The devices that are invariant to body motion ideally assume that every motion that occurs in the patient is the motion due to respiration. On the other hand, the non-invariant systems basically involves only the monitoring of the chest motion from multiple cameras and then obtaining the depth of the region of interest. Some of the devices might also be classified as contact and non-contact devices. An example of a contact based system that is invariant to rigid body motion would be a belt strapped around the chest region \cite{ref3}. Bentur et al. \cite{n1} used contact-based for measuring how children with asthma are responding to medical therapy sessions. Their study underlines that our proposed system could also support medical practitioners or other researchers to monitor therapy response and evolution of the disease. Another widely used commercial device which was introduced by LeoSound, also uses a contact-based method for the respiratory motion measurement during sleep \cite{n2}. Such devices, however, might both produce not completely accurate results since every motion is considered as respiratory motion, while, at the same time having an object strapped around the chest also causes a lot of discomfort, as a result of which it might actually make the person breathe inefficiently. For this reason, we propose to use a non-invariant contactless system. This can be done by obtaining a video using a stereo camera. The solution thereby reduces any of the negative outcomes of the systems that might hinder the breathing motion while at the same time being cheaper and very user friendly, thus enabling it to be used for tracking child respiratory motion at homes on a frequent basis.

Massaroni et al. \cite{n3} proposed a contactless respiratory motion monitoring system based on usage of regular RGB cameras. Even though the system looks effective for homebased usage simply using a webcam, the missing aspect of the 3D reconstruction would highly limit the system to measure 3D motion of the chest. Besides, since the approach works using the color bands of the RGB camera, the system would need white light in the sleeping room to guarantee good results. We believe that our solution works more reliably under low light conditions (because the algorithm doesn't rely on the color components of the images) while providing 3D chest motion which we convert to respiratory signal for detecting abnormalities. 

\section{Methods}
For simple and handy utilization, we have implemented a system that only makes use of a stereo camera. A stereo camera is a passive camera that can be used to produce a depth image of the object under test since it involves two individual cameras. These together provide a distinctive outlook on the same object much as how the human eye works. This significant property of the camera can be used to appropriately determine the depth of an object. The depth of an image as perceived by the camera is directly proportional to the focal length, baseline of the camera and inversely proportional to the disparity. The camera can be used to ideally cover infinite depth, however as the depth increases the error rate increases; thus, maintaining a distance between 0.50 - 1m from the camera produces more accurate results since the breathing motion is more easily visible within this range. At the time of testing, it should be noted that the patient is able to breathe in two different manners: that is either through thoracic breathing or abdominal breathing. A video of the patient breathing should be recorded with a frame rate of 30Hz and a resolution of 480x1280. The sampling rate of the video for \cite{ref4}. Having a sampling frequency as high as 30Hz can result in a large video size, thus halving the sampling rate might be a better solution. Attention should be paid to the extent to which the sampling frequency is lowered, since lowering the sampling rate might affect the quality of data by making it noisier. For specific analysis, a region of interest can also be selected. However, as mentioned earlier, by taking a region of interest, the analysis is being limited to either one of the breathing mechanisms which might thus result in inaccurate results. On the positive side, selecting a region of interest decreases the computation time greatly.

\begin{figure}
  \centering
  \includegraphics[scale=0.65]{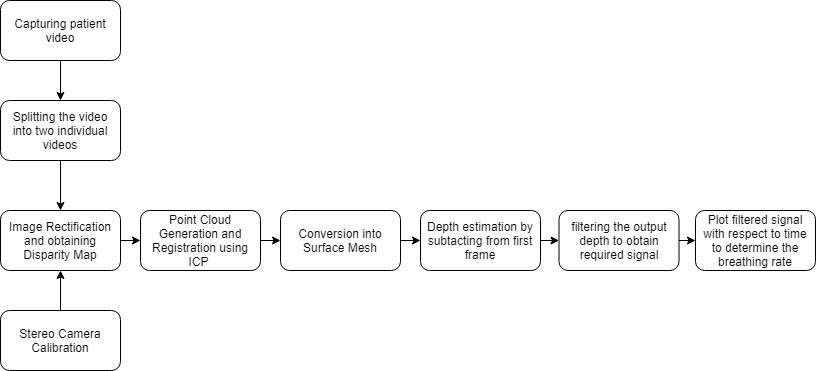}
  \caption{Block Diagram on the process flow of the setup}
  \label{fig:fig1}
\end{figure}
 
\subsection{Stereo Camera Calibration}
Stereo camera calibration: Several images (20 images) of a 7x10 checkerboard with each of the checkerboard squares having dimensions of 10mm are taken from different angles and different aspect points. For the sake of calibration, the Stereo Camera Calibration App of Matlab is used wherein the improper images are directly eliminated thereby reducing the calibration error rate. The stereo parameters obtained are then used for processing the breathing frames.

\subsection{3D reconstruction of chest and tracking respiratory motion}
\begin{figure}
  \centering
  \includegraphics[scale=0.4]{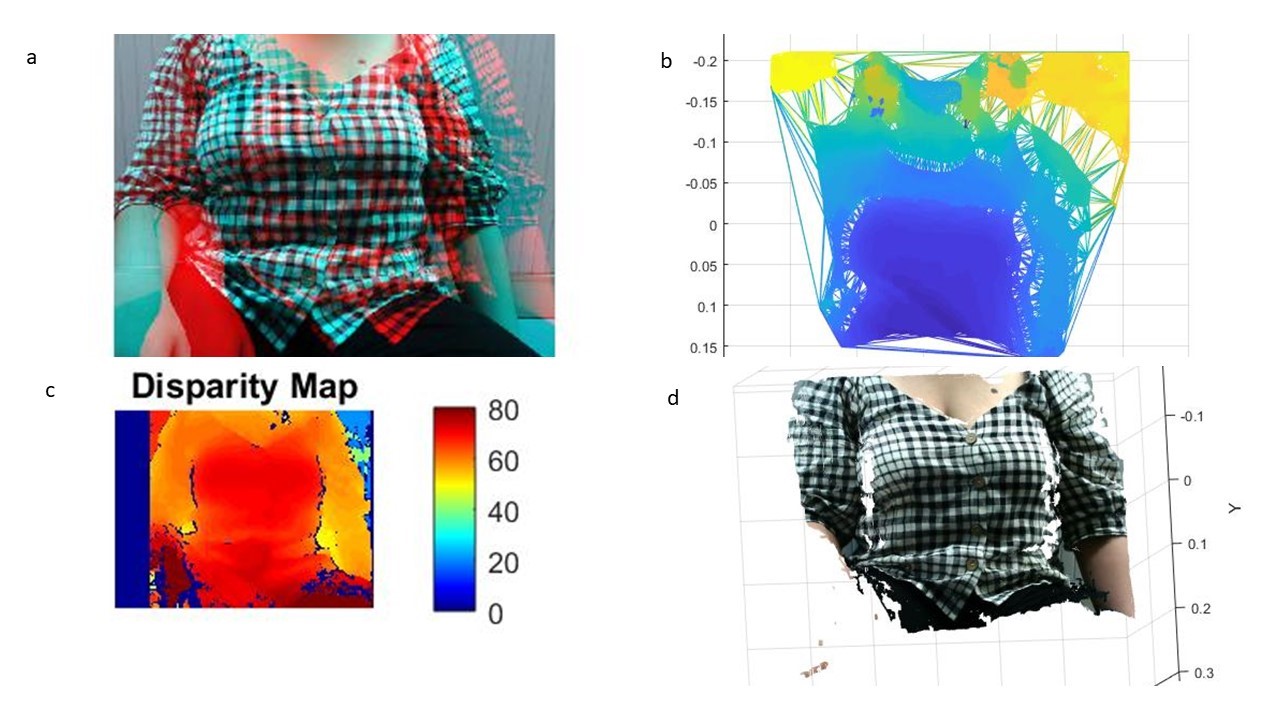}
  \caption{Initial image processing (a) Image Rectification (b) Triangular surface mesh  (c) disparity map (d) Point cloud}
  \label{fig:fig2}
\end{figure}

For the purpose of tracking the respiratory motion of a patient, a video of about 30 seconds to 1 minute is first taken by placing the stereo camera at a fixed distance of about 1m from the patient. Since the obtained video consists of frames from both the cameras of the stereo camera, the video should be split into two individual videos. Each individual video frame then has to be rectified. Image rectification is extremely important in order to reduce the trouble of finding matching points between the frames later on. Upon rectification, it is possible to determine the distance between alike objects in the two images, which in turn helps in computing the disparity. It is very important to set the disparity range correctly in order to obtain a proper output. The disparity data needs to be filtered before it is further converted into a 3D point cloud as can be seen in figure \ref{fig:fig2}. The 3D point cloud is basically a collection of the three-dimensional coordinate system that is obtained from the disparity data in order to produce an accurate 3D digital output of the chest. For further processing of the data, the invalid points that are present in the point cloud need to be eliminated and the point cloud should be de-noised An extremely essential step that should be taken after obtaining the point cloud is aligning all the video frames along the same axis, thereby making the data suitable for tracking respiratory motion. This is done by using the Iterative Closest Point Algorithm \cite{ref5}. The algorithm re-determines the closest point set and continues until it finds the local minimum match between the two surfaces. It works in a few stages, that is, identification of the closest model point for each data point and then finding the least square rigid body transformation relating these point sets. The initial frame is taken as the reference frame for aligning the point clouds to a common axis, thus obtaining the transformation angle. In order to obtain the depth data, the aligned 3D point cloud should then be converted into a surface mesh. The depth information can be acquired by keeping the initial frame as the reference frame (to be taken such that it is the intermediary stage between inhalation and exhalation) and subtracting the depth of every subsequent frame from it. The obtained output will however consist of a lot of noisy data apart from the input signal and needs to be filtered. The noise can be of various types including salt and pepper noise, quantization noise and white noise. This filtered output should then be plotted with respect to time.

Based on the number of breaths that the person takes and the frequency with which it occurs, the person’s breathing condition can be determined. This can be done by determining the number of peaks in the output plot, where in each peak represents a maximum breath that is taken. Based on literature, it is seen that for children below the age of 6, the number of breaths should be between 22-34 breaths per minute while children between 6 to 12 years should be 18-30 breaths per minute \cite{ref6}. If the patient(child) breathes more than 15-17 breaths or less than 10 breaths in 30 seconds a conclusion can be drawn about the breathing condition of the person being tested. A conclusion can be drawn easily if a person is taking more breaths than required that the patient is showing symptoms of asthma.

\section{Experimental setup and results}

The following materials have been used for implementation and testing of this method:

\textbf{Stereo Camera:} Non-Distortion Dual Lens USB3.0 Camera Module Synchronization HD 960P OTG UVC Plug Play Driverless 3D VR Stereo Webcam (~75euro). Output of this camera is stereo images with a resolution of 960x1280 for each separate left and right image.
\textbf{Software:} MATLAB R2019a, including the Image Processing Toolbox.

The accuracy of the system was tested by checking the precision with which the output plot was produced with respect to time and visually matching that output with the input video that was taken of the patient. The input video taken was of a relatively small time frame since it was done in order to test the accuracy of determining respiration. The sampling rate for processing was also maintained at 30Hz. Reducing the sampling rate to 6.7Hz increased the amount of noise that was present in the output. 

After obtaining the initial output, the data needed to be further filtered. This was done by designing a band pass filter with the help of using the Fast Fourier Transform (FFT). From the output of the FFT, the suitable frequency range for only the required breathing data was selected and the rest of the noise was thus ignored. The system was tested by having a volunteer breathe in 4 different manners, including normal breathing, deep breathing, shallow breathing and coughing. For the deep breathing, the volunteer was asked to inhale air very slowly and then again exhale very slowly. On the other hand, the volunteer when asked for shallow breathing was asked to partially inhale and exhale very rapidly. This data served as input into the FFT, thus obtaining a clear output signal from which the breathing pattern can be determined. The number of breaths that are taken are displayed on a dialog box of the user interface (on a pc at the moment) in order to indicate the current health status.

\begin{figure}
  \centering
  \includegraphics[scale=0.45]{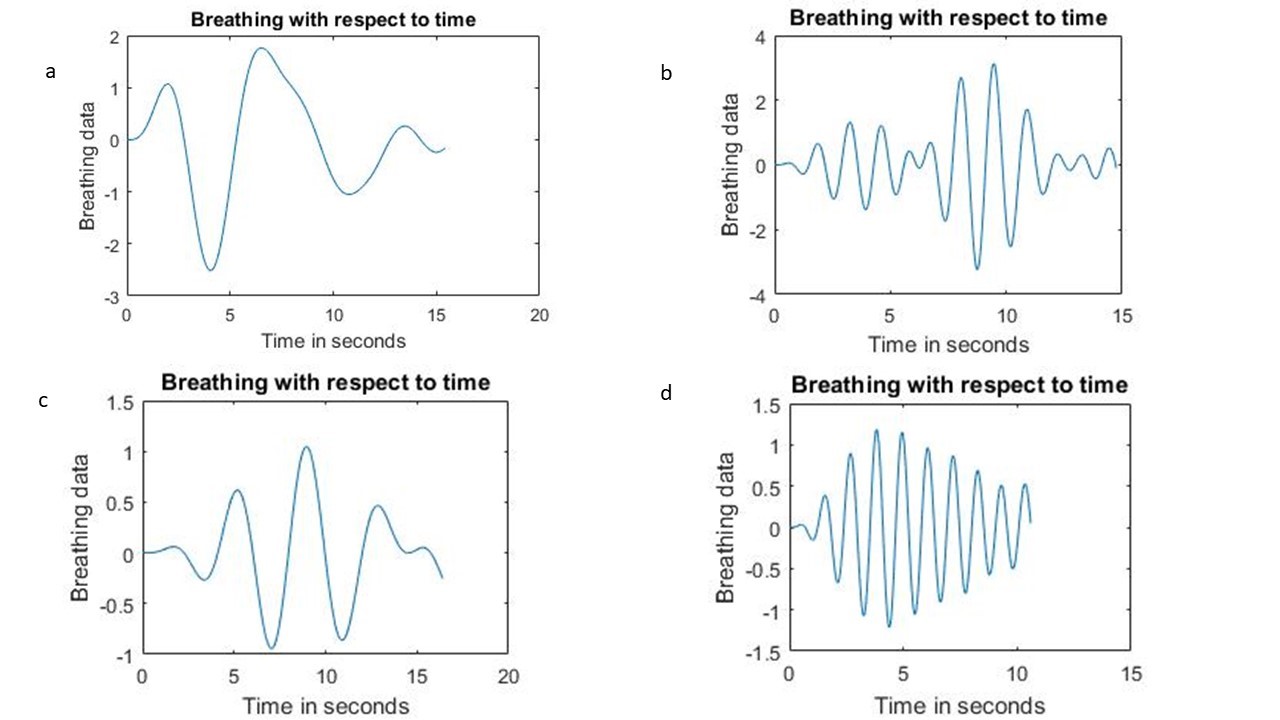}
  \caption{Breathing output for each test (a) long breathing (b) breathing while coughing (c)  breathing while coughing (d) short breathing}
  \label{fig:fig3}
\end{figure}

\begin{figure}
  \centering
  \includegraphics[scale=0.5]{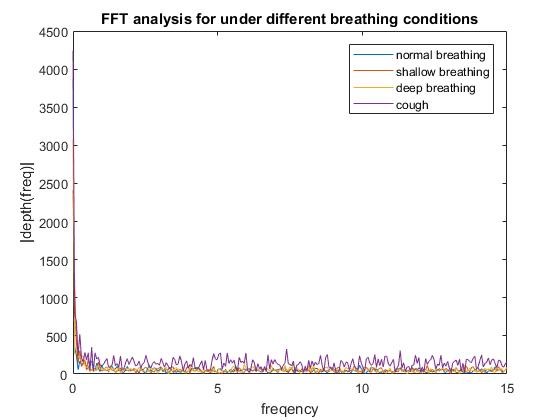}
  \caption{FFT for different breathing input signals}
  \label{fig:fig4}
\end{figure}

\begin{figure}
  \centering
  \includegraphics[scale=0.5]{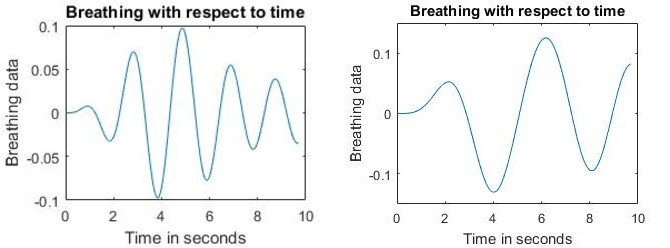}
  \caption{Region of interest. Left: entire chest area. Right: limited region of interest.}
  \label{fig:fig5}
\end{figure}

\begin{figure}
  \centering
  \includegraphics[scale=0.5]{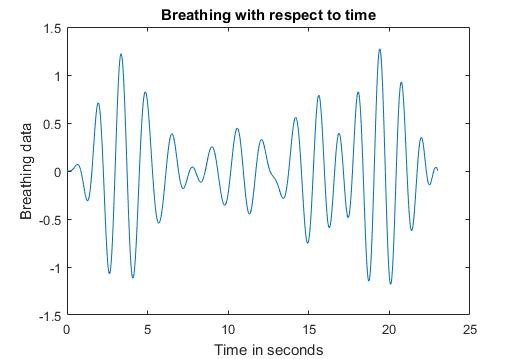}
  \includegraphics[scale=0.5]{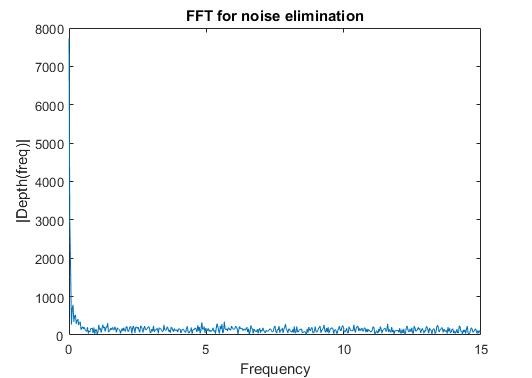}
  \caption{Mixture of breathing input. Left: Output signal. Right: FFT result.}
  \label{fig:fig6}
\end{figure}

However, the frequency range that was selected for every test kept changing and thus could not be generalized to be within a specific range for each different breathing condition as can be seen in figure \ref{fig:fig4}. The filtered output plots of the different breathing conditions can be seen in figure \ref{fig:fig3}. The data for normal breathing was further also evaluated by considering a small region of interest. This saved a lot of computation time, however, the output produced by limiting the size of the region of interest to just a small black point on the chest resulted in a much noisier FFT output response. Thus, the region of interest was changed to the entire chest area. Another issue that also arose was that, the volunteer switched between breathing mechanisms in the middle of the test run as a result of which when only a small region of interest was considered, there were certain cycles that were missed. This can be seen in figure \ref{fig:fig5}. On successful establishment of the previous stages of testing, we performed a test where in a volunteer was asked to shuffle between normal breathing and shallow breathing in order to accurately determine whether the system works efficiently well. The output of the test can be seen in the figure \ref{fig:fig6}.

\begin{figure}
  \centering
  \includegraphics[scale=0.3]{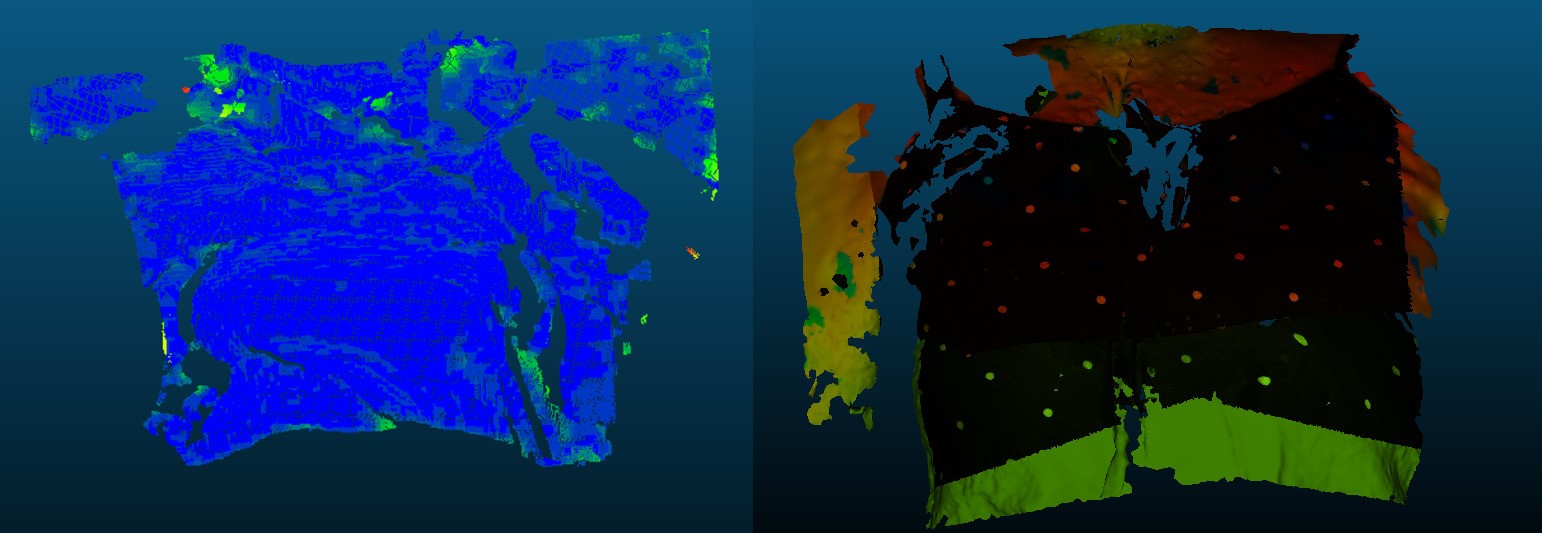}
  \caption{Depth comparison. Left: 3D reconstruction with stereo camera. Right: 3D reconstruction with 3dMD.}
  \label{fig:fig7}
\end{figure}

In order to conclusively determine whether the designed system was successful, we compared our output with that of the more expensive and precise medical system, the 3dMD. This was done by calculating the maximum expansion of the chest at the time of deep breathing from the intermediary state between inhalation and exhalation. Due to the extreme complexity of doing this comparison in MatLab, the point clouds from the 3dMD and the stereo camera were entered into point cloud processing software called as CloudCompare \cite{ref7}. The maximum depth of the 3dMD was found to be 36.0661mm with an error rate of 1.783 while our alternative had a depth of 38.4005mm and an error rate of 1.542. In Figure \ref{fig:fig7} we can see the output image from CloudCompare for both the 3dMD and the stereo camera.

A simple low cost system that is easy to use and comprehend was thus developed. However, there is still scope for more research when it comes to elimination of the noise, since the frequency range that could be used for determining the breathing pattern of the patient kept changing with every different test that was performed. The difference in the measurement that arose between the depth estimation of the 3dMD system and the Stereo camera could be due to either the difference in the induced noise(that could be due to the difference in the relative thickness of the clothes) or it could also be due to the camera calibration. The system should also be generalized to be independent of whether the first frame is the intermediate breathing position or not. This could be done by taking a cycle of breathing before starting the data and further storing the reference position from that. The setup could also not be tested with children, but that’s not a major problem since it can be solved easily by adjusting the maximum and minimum number of breaths that can be taken by the children in 30 seconds.  

\section{Conclusions and future work}
The output performance of this system was verified by testing under several breathing conditions and was also compared with the benchmark system that is 3dMD, that produced a small depth difference of about 2mm. This system can be incorporated without hindering the persons breathing, in a completely non-contact manner. It can thus successfully be used to monitor the patients breathing on a regular basis. This is especially required for children since after a certain age the child that used to have asthmatic symptoms might not necessarily continue to suffer from the same. Thus, parents can keep a check of their child’s breathing and on unfortunate events can further go on to seeing a doctor. Conclusively, with a little bit more research, this will be a very successful method for the detection of respiratory disorders like asthma.

\bibliographystyle{unsrt}  
\bibliography{main}  

\end{document}